\documentstyle[aps,prb,multicol,epsfig]{revtex}
\tighten
\newcommand{\beq}{\begin{equation}}
\newcommand{\eeq}{\end{equation}}
\newcommand{\bea}{\begin{eqnarray}}
\newcommand{\eea}{\end{eqnarray}}

\begin{document}
\title{ Dynamical Mean Field Theory and Electronic Structure Calculations}
\author{R. Chitra and G. Kotliar }
\address
{ Center for Materials Theory, Department of Physics and Astronomy, Rutgers University,
Piscataway, NJ 08854, USA}
\date{\today}
\maketitle
\begin{abstract}

We formulate the dynamical mean field theory directly
in the continuum. For a given definition of the local
Green's function, we show the existence of a unique functional,
whose  stationary point  gives the physical 
local Green's function of the solid.
We present the diagrammatic rules to
calculate it  perturbatively  in the interaction.
Inspired by the success of dynamical mean field calculations
for model Hamiltonian systems,
we present approximations to the exact saddle point
equations which may be used in the  strong coupling regime, 
by using mappings onto
generalized quantum impurity models. 

\end{abstract}
\section{Introduction }

The calculation of the electronic structure of solids starting from
first principles has been the subject of intensive investigations in
the past.  The most widely used method is the density functional
theory (DFT), used  frequently in the local  density approximation (for a recent
review  see Ref.\onlinecite{kohn}). 
In principle, this method is good for calculating   
ground state properties only. Time dependent extensions of
 density functional
theory \cite{gross}  have also been formulated and  allow the calculation of density and
spin correlation response functions which contain information about
excited state properties.
This method has been very successful in  understanding 
weakly   correlated solids but has serious drawbacks when dealing with strongly
correlated systems. 
A second general formulation of the quantum many body
problem 
is the Green's function method as formulated early on by Luttinger and
Ward \cite{ward}  and Baym  and Kadanoff \cite{baym}. 
In this method one sets up fully self consistent 
equations for the one particle Green's function using a functional i.e.,
an effective action for the Green's function,
which has a well defined diagrammatic interpretation in terms of skeleton graphs.
However,  when considering a realistic band structure,
this method becomes  fairly intractable  beyond the  lowest
order or GW approximation \cite{gw},  and is inadequate for 
  strongly  correlated problems.

The purpose of this paper is to describe the foundation of a third
approach, namely, the dynamical mean field method (DMFT) in the continuum,
in relation to the two
other techniques described above.  
Our main motivation
stems from the success of 
 DMFT in dealing with  model Hamiltonians of  strongly correlated
 electrons on a lattice\cite{review}.
In fact, there are already several
works attempting implementations of dynamical mean field ideas
to electronic structure calculations \cite{lichtenstein}.
This paper is  an attempt to provide a theoretical underpinning
to these computational efforts. 

The local spectral function is a central concept in dynamical
mean field theory. 
The  goal of the theory is to compute the one electron addition
and removal spectra, and quantities which can be naturally
expressed in terms of the local spectra. One can view the local
spectral function as providing a frequency resolution of the local
density.  The hope is that by using a more complicated quantity
to formulate the theory, one may be able to formulate flexible
approximations, which work well even in strongly correlated
situations, where density functional theory
in the local approximation is known to fail.
Indeed, recent work on model calculations has shown that
DMFT plays a very  useful role 
in situations where significant changes in the local
spectral function take place. These rearrangements of 
one particle spectral weight  may not be accompanied by significant
changes in the electron density, and may not be captured well
in approximate implementations of density functional theory. 
This occurs in  metal insulator transitions which are
accompanied by very small symmetry changes or, small
structural charges.
Examples of this kind of transition are 
the paramagnetic insulator to paramagnetic metal transition
in $V_2 O_3$ and $Ni Se_{1-x} S_x$ , the $\alpha$ to $\gamma $ transition
in cerium,
the transitions form the   $\beta $ to  the $\delta$ phase in  plutonium 
or the more recently observed metal insulator transition
in  $Y_{1-x}Ti 0_3$ \cite{iga}. 
We also expect the dynamical mean field theory
to provide useful descriptions of phases where local moments are well
developed but not ordered such as in $\delta $ plutonium \cite{savrasov}.

In this paper, we will show that 
in  general DMFT functionals
are more cumbersome in their diagrammatic formulation 
than  the Ward Luttinger  functional for the full Greens function. 
Nevertheless, our motivation for presenting this formulation here
is that DMFT 
 is
very  promising in terms of
practical implementation  in very strongly correlated situations. 
The reason for
that is that it can be formulated in terms of impurity models  embedded
in a medium whose properties have to be evaluated self consistently
\cite{gk}.
A
large number of numerical methods for treating the resulting 
dynamical mean field   equations
have been developed in the context of model Hamiltonian calculations 
\cite{review} and  these techniques can and should be extended
to more realistic models of solids.

Conceptually, the three techniques are very similar: one defines a
functional $\Gamma$ of the quantity of interest X (X being the density in the
DFT  case, the  local Green's function (to be defined later)
 for DMFT or the full  Green's function in the
Ward Luttinger  technique) such that  the  
 minimization of $\Gamma$  selects the physical value of the quantity of
interest.  In all the  techniques, there is an exact functional $\Gamma _{exact}$
 having the
required property but its explicit analytic  form is  very difficult to determine
.  In all the three cases,   approximations to  $\Gamma _{exact}$  are necessary to
 obtain practical results.  We adopt a fairly general formulation to set up a  diagrammatic
 expansion for the exact functional
of the local Green's function, whose saddle point gives  the physical local Green's function.
These are presented  in the first  part of this paper and 
 some possible  approximate implementations are presented in the second part of this paper. 

Extending the more standard  lattice model 
calculations,
our work 
provides a foundation for  doing ab-initio  dynamical mean field theory plus
 electronic structure calculations in the
continuum. 
The goal  of our  work is twofold.  First we want to establish the
exact diagrammatic   content of $\Gamma $ of the dynamical mean field
formalism for  realistic models of solids where electrons move in
the periodic potential of a crystal, as a formal framework for
performing electronic structure calculations or, in other words,
to construct a functional for which a dynamical mean field approximation
gives the exact answer to the electronic structure problem. This is the
content of Sec. II. 
This perspective should be contrasted with an alternative
formulation  \cite{chitra2} which will be presented in a
separate publication, where dynamical mean field theory
will be viewed as an {\it approximate} solution to the 
electronic structure calculation problem. 
Secondly, in Sec. III,
encouraged by the success of 
dynamical mean field theory in the treatment of model hamiltonians \cite{review},
 we  suggest some  useful approximations for this
functional.  
This work should provide  a conceptual framework for  the recent
efforts in combining realistic density functional calculations with
techniques for solving impurity models in a self consistent environment
\cite{lichtenstein}. 
Finally, we notice that our work can be regarded as a generalization
of Ref.\onlinecite{gabi}, where lattice DMFT was given a functional formulation
in terms of a field which is conjugate to the local spectral function.
This formulation has given useful insights into the physics of the
Mott transition.

\section{ \bf Effective Action Formalism}

In this section, we develop the 
effective action formalism for DMFT.   
This is in the same spirit as the effective action approach to
Kohn-Sham density functional theory (where the bilinear operator is just
the electron density) which was developed in a series of publications by Fukuda {it et al}\cite{fukuda}  and Valiev and Fernando \cite{fernando}. 
We first set  the notation. 
The continuum is divided into unit cells labeled by ${\bf R}$ and ${\bf r}$ 
is the vector
defined within a unit cell. 
A general "local" bilinear operator  has the form 
\beq
O^{\alpha \beta}({\bf r},{\bf r}^{\prime},\tau,\tau^{\prime}) = 
\int{d{\bf  x} d{\bf y}} 
K({\bf r},{\bf r}^\prime\vert {\bf x}{\bf y}) \psi^+_{\alpha}({\bf x}, \tau) \psi_{\beta}(
{\bf y},\tau^{\prime})
\label{oper}
\eeq
\noindent
where $\psi^+$ and $\psi$ are the electron creation and annihilation
operators and 
 $K$ is a kernel to be specified below.  ${\bf x},\tau$ are the space and imaginary
time coordinates and $\alpha,\beta$ denote the spin indices. With the definition
 $\langle T_\tau O\rangle= A({\bf r},{\bf r}^\prime,\tau,\tau^{\prime}) $
(\ref{oper}) leads to 
\beq
A^{\alpha \beta}({\bf r},{\bf r}^{\prime},\tau,\tau^{\prime}) =
\int{d{\bf x} d{\bf y}}
K({\bf r},{\bf r}^\prime\vert {\bf x}{\bf y}) G^{\alpha \beta}({\bf x},{\bf y}, \tau,\tau^{\prime})
\label{green}
\eeq
\noindent
In terms of the cell indices and unit vectors   the space coordinates read 
  ${\bf x}={\bf R}+{\bf u}$ and ${\bf y}={\bf R}^\prime +{\bf u}^\prime$.  
Here, ${\bf u}$, ${\bf r}^\prime$, ${\bf r}$${\bf u}^\prime$ are vectors inside
a unit cell. 
The kernel  $K$  projects the full $G$ onto a
"local" Green's function $A$ defined within a unit cell.  By choosing the kernel $K$ to be sufficiently localized around ${\bf R} -{\bf R}^\prime \approx 0 $, we 
extract or project out local information from the full Greens function.
 There is a great deal of flexibility in deciding what one would
like to define  as  the "local"
single particle Green's functions.  
The  simplest choice  would be  
\beq
K= \delta_{{\bf R},{\bf R}^\prime} \delta({\bf r} - {\bf r}^\prime) \delta({\bf r} - {\bf u})  \delta({\bf r}^\prime - {\bf u}^\prime) 
\label{kernela}
\eeq
which gives a resolution of the density with frequency
\beq
\rho(r) = {1 \over \beta}\sum_{n} \sum_{\alpha} \exp^{i\omega_n 0^+} {\bar A}^{\alpha \alpha}(i\omega_n, r,r)
\eeq
where ${\bar A}(i\omega_n, r,r)$ is the spectral function corresponding to A, $\omega_n$
are the Matsubara frequencies and $\beta$ is  the inverse temperature. 
Experience has suggested however, that
it may be useful to retain local phase information among different
orbitals in the local Greens functions.  The definition 

\beq
K= \delta_{{\bf R},{\bf R}^\prime} \delta({\bf r} - {\bf u})
  \delta({\bf r}^\prime - {\bf u}^\prime) 
\label{kernel1}
\eeq
\noindent
gives us simply the restriction of the full Green's function to  a given 
unit cell.
For  ${\bf r}={\bf r}^\prime$, we regain the former choice of
 local spectral function. 
Note that (\ref{kernel1}) and hence the corresponding Green's function $A$ is not  invariant under simultaneous
lattice translations of 
 $ {\bf r}  $ and ${\bf r}^\prime$ now viewed 
as coordinates defined throughout  the whole crystal.
A  kernel with translation invariance is  obtained by considering  
\beq
\label{trans}
A^{\alpha \beta}({\bf r},{\bf r}^{\prime},\tau,\tau^{\prime}) = \sum_{{\bf G},{\bf G}^\prime}\sum_{\bf k } 
G^{\alpha \beta} ({\bf k}+{\bf G},{\bf k} +{\bf G}^\prime ,\tau,\tau^\prime)
 \exp{i{\bf G}.{\bf r} -i{\bf G}^\prime.{\bf r}^\prime}
\eeq
\noindent
Here the ${\bf G}$ are the reciprocal lattice vectors and ${\bf k}$ runs over the Brillouin
zone.  Rewriting the   right hand side of (\ref{trans})
in real space coordinates  and summing over the reciprocal lattice
vectors, we obtain a kernel invariant under lattice translations.
\beq
K=\sum_{\bf k} \exp{i{\bf k}.({\bf x}-{\bf y})}\delta({\bf r} - {\bf u})
  \delta({\bf r}^\prime - {\bf u}^\prime)                            
\label{choice2}
\eeq
\noindent
where ${\bf k}$ runs over the Brillouin
zone. 
Finally,  given a set of tight binding orbitals, there is a natural kernel
 $K$, associated with  this set.
The 
 full Green's function is expressed in  a set of tight binding  orbitals
$\phi_{\it l}$ as follows:
\begin{equation}
\label{decom}
G^{\alpha \beta}({\bf x},{\bf y})  = {\sum_{{\it lm}{\bf R}_1 {\bf R}_2} }\phi_{\it l}({\bf x} - {\bf R}_1)
\phi_{\it m}({\bf y} - {\bf R}_2)  G^{\alpha \beta}_{\it lm}({\bf R}_1,{\bf R}_2)
\end{equation}
Similarly, the local Green's function can be projected onto this set
as    
\begin{equation}
\label{deloc}
A^{\alpha \beta}({\bf r},{\bf r}^\prime)  = \sum_{{\it lm}{\bf R}_1}  \phi_{\it l}({\bf r} - {\bf R}_1)
\phi_{\it m}({\bf r}^\prime - {\bf R}_1)  G^{\alpha \beta}_{\it lm}({\bf R}_1,{\bf R}_1
)
\end{equation}
Using (\ref{decom}), (\ref{deloc}) can be rewritten as
\begin{equation}
\label{del}
A^{\alpha \beta}({\bf r},{\bf r}^\prime)  = \sum_{{\it lm}{\bf R}_1 {\bf R}_2} \int_{{\bf x},{\bf y}}\phi_{\it l}({\bf x} - {\bf 
R}_1) \phi_{\it l}({\bf r} - {\bf R}_1)
\phi_{\it m}({\bf y} - {\bf R}_2) \phi_{\it m}({\bf r}^\prime - {\bf R}_2)
G^{\alpha \beta}_{\it lm}({\bf x},{\bf y})
\end{equation}
This leads to the following orbital decomposition for the kernel 
\begin{equation}
\label{kerorb}
K({\bf r},{\bf r}^\prime\vert {\bf x}{\bf y}) =\sum_{{\it lm}{\bf R}_1 {\bf R}_2} \phi_{\it l}({\bf x} - {\bf
R}_1) \phi_{\it l}({\bf r} - {\bf R}_1)
\phi_{\it m}({\bf y} - {\bf R}_2) \phi_{\it m}({\bf r}^\prime - {\bf R}_2)
\end{equation}
\noindent
{}From these we see that different choices of the basis functions lead to
different kernels corresponding to different definitions of the "local"
Green's function.
We mention that  recent works \cite{lichtenstein} incorporating DMFT concepts in electronic
structure calculations have all   used the  LMTO \cite{andersen} basis set.

We now use the formalism  developed in  \cite{fukuda}  to construct the
effective action $\Gamma$ for  the
local Green's function  defined by (\ref{green}).
These methods prescribe a scheme to construct
an analog of 
 the "Kohn-Sham" correlation potential  calculated in the conventional
density functional theory \cite{fernando}.
 The idea is to construct a perturbative expansion in the interaction strength,
for the effective action  $\Gamma$.
 We use the imaginary time
formalism in the rest of the paper. 
 The partition
function $Z$ for a system of interacting electrons    coupled to a source
field $J$ via the operator $O$ is  
\bea
Z[J^{\alpha\beta}]&=&\exp W[J^{\alpha\beta}] \nonumber \\
&=& 
  \int D\psi D\psi^{+}  \exp [{-S + \int J^{\alpha \beta}({\bf r},{\bf r}^{\prime},\tau,\tau^{\prime})
O^{\alpha \beta}({\bf r},{\bf r}^\prime , \tau , \tau^{\prime}) }]
\label{part}
\eea
\noindent
The coordinates ${\bf x}$ run over the entire range of the system and 
It is crucial to note that the indices $\alpha \beta$ are not summed over in the
second term in the exponential in (\ref{part}).
The quantity $W[J]$  introduced above is
 the generator of all connected
diagrams. 
 $S$ is  the action of a system of
electrons with Coulomb interactions between them.
\bea
S&=& \int dx\sum_{\alpha}  \psi^{+}_{\alpha}(x) [ \partial_{\tau}- {{\bigtriangledown^2} \over
{2m}} +V_{X}(x)]\psi_{\alpha}(x) \\
& + &
 {{e^2} \over 2} \sum_{\alpha \beta}\int dx dx^{\prime} \psi^{+}_{\alpha}(x)
 \psi^{+}_{\beta}(x^\prime) U_c(x-x^{\prime})
 \psi_{\beta}(x^\prime)
 \psi_{\alpha}(x)
\label{iham}
\eea
\noindent
In (\ref{iham}), $x=({\bf x},\tau)$,  
 $V_X$ represents the  ionic potential 
the electron moves in and the Coulomb interaction is
$U_c(x-x^\prime)= \vert {\bf  x} - {\bf x^\prime} \vert ^{-1} \delta(\tau-\tau^
{\prime})$.
The effective action  $\Gamma$ is defined by  the Legendre transform  
\beq
\Gamma[A^{\alpha \beta}] = W[J_{\alpha \beta}] - \int_{{\bf r},{\bf r}^\prime} J^{\alpha \beta }(r,r^{\prime}) A^{\alpha \beta }(r,r^{\prime})
\label{gamma}
\eeq
\noindent
A naive expectation might be that since $A$ is just a "local" Green's
function, the effective action functional $\Gamma$ is given by the
usual Baym-Kadanoff functional. Here we find that this is not the
case, rather the restriction of ${\bf r},{\bf r}^\prime$ to a unit cell induces
non-trivial corrections to the $\Gamma$ as will be illustrated  
below.
To obtain $\Gamma$ as a functional of $A $,
 the $J$ dependent terms in (\ref{gamma})  should be rewritten in
terms of  $A$.   Since this is not tractable in the
presence of interactions, we use the 
 inversion method \cite{fukuda} which 
 involves an expansion of $J,W$ 
in terms of the coupling constant of the interaction (for example, $e^2$ for
coulomb interactions)
 and $J_0$, where $J_0$ is the source
in the absence of the interaction. The crucial ingredient of this
method is that  $J_0$ is chosen
so as to reproduce the correct  local Green's function $A^{\alpha \beta}$ of the 
{\it full interacting
problem} i.e.,  
\beq
{{\delta W_0[J_0^{\alpha \beta}]} \over  {\delta J_0^{\alpha \beta}}}  = A^{\alpha \beta}
\label{joa}
\eeq
\noindent
where $W_0$ is the non-interacting free energy.
Using (\ref{joa}), the series expansion for the $W[J_0^{\alpha \beta}]$ and $J[J_0^{\alpha \beta}]$ can 
and rewritten in terms of $A$ and thereby  resulting in a series  for
the effective action $\Gamma[A^{\alpha \beta}]$
\beq
\Gamma[A^{\alpha \beta}] = \sum_{i=0}^\infty  e^{2i} \Gamma_i[A^{\alpha \beta}]
\label{gamser}
\eeq
\noindent
The  exact local function $A$ of the interacting system is obtained by minimizing (\ref{gamser})
\beq
{{\delta \Gamma[A^{\alpha \beta}]} \over {\delta A^{\alpha \beta}}} =0
\label{var}
\eeq

\subsection{ Diagrammatic rules for $\Gamma$}

The first step is the construction of the  non-interacting
"Kohn Sham" Green's
function $G_{0}[J_0]$ 
\beq
G_{0}^{\alpha \beta} (R+r ,R^\prime +r ^\prime) = \left[\left(\partial_\tau-
{{\bigtriangledown^{2}}\over {2m}} +
v_{X}({\bf R}+{\bf r})\right)\delta_{{\bf R} {\bf R}^\prime}
\delta(r-r^\prime)\delta_{\alpha \beta} -
 J_0^{\alpha \beta}(r,r^\prime) 
\delta_{{\bf R},{\bf R}^\prime}
\right]^{-1}
\label{ksgr}
\eeq
\noindent
Note that in the above expression  $R+r\equiv({\bf R}+{\bf r},\tau_r)$ and
equivalently, $r={\bf r},\tau_r$.
 $J_0^{\alpha\beta}$ is the Kohn-Sham hybridization  function defined earlier.
$J_0$ is chosen  such that
$G_{0}$ gives the prescribed local Greens function
$A
 $  of the
full interacting problem.
Using $G_0$,  the Coulomb interaction lines ~$e^2 U_c
(R-R^\prime+r-r^{\prime})$ and a propagator ${\bf D}$ to be defined below  and
following the method of \cite{fukuda,fernando}, we can  
 obtain 
$\Gamma$ to all orders in the coupling $e^2$.
The zeroth order term  is $W_0= 
{\rm Tr} \log G_0$ and the first order term   is given by the sum of
the Hartree and Fock diagrams shown in Fig.1.  The higher order  terms  which form the
Kohn-Sham exchange-correlation potential  in the conventional density
functional theory are constructed
according
to the following diagrammatic rules.
At every order : 

\noindent
1.  Draw all connected diagrams made of Kohn-Sham propagators
$G_0$ and Coulomb interaction lines  $e^2 U_c(R-R^\prime +r -r^\prime)$ 
 with the corresponding weight factors. \\

\noindent
 2.  Eliminate all  graphs  which are one vertex irreducible i.e.,
 separable  by cutting a single
Coulomb interaction line.
 
\noindent
 3.  For each two-particle reducible (2PR) graph (i.e. any graph that
can be separated by cutting two propagator lines),  perform the
following operations.$^{5}$ \\ 
\smallskip

\indent (a) Separate the graph by cutting  the 2PR propagators.\\

\indent (b) For each of the two resulting graphs join two external propagators.\\

\indent (c) Connect the two graphs via the inverse two particle  propagator
${\bf D}$
which is the inverse of the connected correlation function $\Pi$  
\begin{eqnarray}
\nonumber
\Pi_{\mu\nu \delta \eta} (r,r^\prime,u,u^\prime)  &= &
\sum_{{\bf R},{\bf R}^{\prime},{\bf U}, {\bf U}^\prime} 
\delta_{{\bf R},{\bf R}^\prime} 
\delta_{{\bf U},{\bf U}^\prime}
\langle \psi^{+}_{\mu}(R+r)\psi_{\nu}(R^{\prime}
+r^{\prime}) \psi^+_{\delta}(U +u)
\psi_{\eta}(U^\prime +u^\prime) \rangle \\
&=& 
-\sum_{{\bf R},{\bf U}} 
G_0^{\mu \eta}(R+r,U +u^
{\prime}) G_0^{\delta \nu}(R+r^\prime,U +u)
\end{eqnarray}
\noindent
In the above expressions  $U+u\equiv({\bf U}+{\bf u},\tau_u)$. 
The inverse propagator $\bf D$  appears in the diagrammatics of
the inversion method,  as a result of    
  corrections to $J_0$ from the interaction which are
constrained so that
one reproduces the 
desired local Green's function $A$ order by order in perturbation theory.

\indent (d) Repeat the procedure until no new graph is produced.\\

\indent (e) Sum up all the resulting graphs including the original
graph.\\

\begin{figure}
\centerline{\epsfig{file=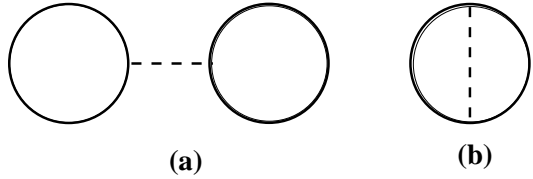,angle=360, width=7cm}}
\caption{a) Hartree and b) Fock diagrams}
\label{fig}
\end{figure}
In Fig.2, we  show the  second order diagrams which contribute  to $\Gamma$.
Here the full lines denote particle propagators $G_0$ and the dashed line is
the Coulomb vertex. 
 Figs. 2a and 2b are examples of diagrams 
 which  are both   one vertex and $2$ particle irreducible. 
\begin{figure}
\centerline{\epsfig{file=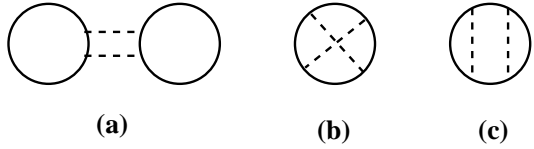,angle=360, width=7cm}}
\caption{ Diagrams at second order}
\label{fig1}
\end{figure}
\noindent
However,  since the diagram in  Fig. 2c  is  two particle  reducible,  it 
transforms   according to Rule 3 as shown in Fig. 3.  The thick double line denotes the two particle
inverse propagator ${\bf D}$.
\begin{figure}
\centerline{\epsfig{file=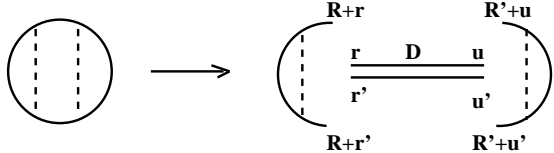,angle=360, width=7cm}}
\caption{ Dressing of a two particle reducible diagram}
\label{fig2}
\end{figure}
\subsection{ Self-consistency }

The self-consistent procedure for obtaining the full $A$ is of the same
kind as that prescribed for $\Gamma[\rho]$ in Ref.\onlinecite{fernando}.
\begin{itemize}

\item Choose a $J_0^{\alpha \beta}(r,r^\prime)$.

\item  Substitute this in (\ref{ksgr}) to
obtain the  Green's function $G_0^{\alpha \beta}(R + r, R^\prime + r^\prime
 )$.

\item Use this $G_0^{\alpha \beta}$ to evaluate the series for
 $\Gamma\equiv \Gamma_0 +\Gamma_H +\Gamma_{xc}$.
Here $\Gamma_H$ denotes the Hartree contribution and $\Gamma_{xc}$ represents
the contribution of the Fock and all higher order diagrams to $\Gamma$.
Determine the new "Kohn-Sham" $J_0^{\alpha \beta}$ defined as
\beq
J_0^{\alpha \beta} = -\left[{{\delta \Gamma_H} \over {\delta A^{\alpha \beta}}} +  
 {{\delta \Gamma_{xc}} \over {\delta A^{\alpha \beta}}}  \right]
\label{j0sc}
\eeq

\item
Using the $J_0$ determined by (\ref{j0sc}), go back to step 1 and iterate until
self-consistency is achieved.
\end{itemize}

\noindent
In density functional theory, the second term in (\ref{j0sc}) is called the
exchange-correlation potential $V_{xc}$.   
	The  scheme  presented above, hinges on  the fact that 
the Legendre transform defined in (\ref{gamma}) is invertible.
Mathematically,  neglecting the spin indices, the transformation is invertible if
${{\delta^2 \Gamma} / { \delta A^2}}$
has no zero eigenvalues.
  Within  the inversion method where $\Gamma$ is evaluated perturbatively in
the interaction strength, the invertibility condition described above
reduces to
the following  equation :
\beq
( {{\delta W_0[J_0]} \over {\delta J_0}} - A)
 {{\delta J_0} \over
{\delta A}}=0 \label{invp} \eeq This yields (\ref{joa}) provided  ${{\delta J_0}/ {\delta A}}$ has no zero eigenvalues or equivalently, \beq {{\delta^2 \Gamma_0} \over { \delta A ^2}} \ne 0 \eeq Typically these transformations are invertible for generic values of the variable $A$. 
Non invertibility 
 is related to a bifurcation of the stationary solution of the functional
$\Gamma$ and  
signals  a phase transition.
This was exploited in Ref.\onlinecite{gabi}, to analyze  the
energetics of a system near a Mott transition.  The functional
constructed here can be used in conjunction with the techniques
of Ref.\onlinecite{gabi}, to  provide more realistic
descriptions of electronic phase transitions.  
Finally, we mention that we have not yet  addressed rigorously, 
 the  questions corresponding  to the issues of  $v$ and $N$ representabilities
 in  density functional theory  \cite{kohn} i.e., what is  the domain of
definition of our functional or equivalently, what is the range of the  function $A$?.
 The formal construction presented here is valid for $A$'s which are Kohn Sham
 representable, and we know that the physical Green's function is Kohn Sham
 representable, with $J_0$  being the physical  "local" self energy.

\subsection{ Example}

For the sake of clarity,  we consider the case of spinless electrons with non-local
interactions.
In this case,
\begin{equation}
\Pi(r,r^\prime,u,u^\prime)  =
-\sum_{{\bf R},{\bf U}} 
G_0(R+r,U +u^
{\prime}) G_0(R+r^\prime,U +u)
\end{equation}
\noindent
 We present the various terms
upto second order in $e^2$,
 contributing to
  $\Gamma[A]$ for the kernel of (\ref{kernel1}).
The zeroth order term is  
$\Gamma_0[A]= {\rm Tr} \log G_0 -J_0 A$. At first order, we have the contributions
from the Hartree and Fock diagrams 
 shown in Figs. 1a and 1b.
\beq
\Gamma_1^H
  = \int dr dr^{\prime} \sum_{{\bf R},{\bf R}^\prime}
G_0(R+r,R+r) G_0(R^\prime +r^\prime,R^\prime+
r^\prime) 
      U_c(R-R^\prime+r-r^\prime)
\eeq
\noindent
The contribution from the Fock term is as follows:
\beq
\Gamma_1^F =
  \int dr dr^{\prime} \sum_{{\bf R},{\bf R}^\prime}
G_0(R+r,R^\prime+r^\prime) G_0(R^\prime +r^\prime,R+r)
      U_c(R-R^\prime+r-r^\prime)
\eeq
Similarly the diagrams which contribute to second order are
shown in Figs.(2a-2c).
\bea
\nonumber
\Gamma_2^a& =&  \int dr dr^{\prime} du du^{\prime} \sum_{{\bf R},{\bf R}^{\prime}
{\bf R}_1,{\bf R}_1^{\prime}} G_0(R+r,R_1+u) G_0(R_1+u,R+r) \\
      && U_c(R-R^\prime+r-r^\prime) G_0(R^\prime +r^\prime, R_1^{\prime}+
u^{\prime})
      U_c(R_1-R_1^\prime+u-u^\prime) 
      G_0(R_1^\prime +u^\prime, R^{\prime}+ r^{\prime}) \\
\nonumber
\Gamma_2^b  &=&
  \int dr dr^{\prime} du du^{\prime} \sum_{{\bf R},{\bf R}^{\prime}
{\bf R}_1,{\bf R}_1^{\prime}} G_0(R+r,R_1+u) G_0(R_1+u,R^\prime+r^\prime) 
       U_c(R-R^\prime+r-r^\prime)\\ 
&& G_0(R^\prime +r^\prime, R_1^{\prime}+
u^{\prime})
      U_c(R_1-R_1^\prime+u-u^\prime) 
      G_0(R_1^\prime +u^\prime, R+ r) \\
\eea
\noindent
\bea
\nonumber
\Gamma_2^c 
& =&  \int dr dr^{\prime} du du^{\prime} \sum_{{\bf R},{\bf R}^{\prime}
{\bf R}_1,{\bf R}_1^{\prime}} G_0(R+r,R_1+u)  G_0(R_1+u,R_1^{\prime}+u^{\prime}) \\
      && U_c(R-R^\prime+r-r^\prime) G_0(R_1^\prime +u^\prime, R^{\prime}+
r^{\prime})
      U_c(R_1-R_1^\prime+u-u^\prime) 
      G_0( R^{\prime}+ r^{\prime},R+r)\\
\label{2pr}
\eea
\noindent
Since Fig.2c is 2PR,  
following the prescription (3a-3c)
presented earlier, we  subtract the
following term from the 2PR diagram of (\ref{2pr}) 
\bea
\nonumber
\Gamma_2^{cr}&=&  
 \int \sum_{{\bf R},{\bf R}^{\prime},{\bf R}_1,{\bf R}_1^\prime,{\bf R}_2,{\bf R}_2^\prime}
G_0(R_1+r_1,R_1^\prime+r_1^\prime)
G_0(R_1+r_1,R+r) G_0(R_1^\prime+r_1^\prime ,R+r^{\prime})\Pi^{-1}(r,r^{\prime},u,u^{\prime})\\
&&  G_0(R^{\prime}+u,R_2+r_2)
G_0(R_2^\prime+r_2^\prime, R^{\prime}+u^{\prime})
G_0(R_2+r_2, R_2^\prime+r_2^\prime ) U_c(R_1-R_1^\prime+r_1-+r_1^\prime
) U_c(R_2-R_2^\prime+r_2-+r_2^\prime)
\label{2prc}
\eea
\noindent 
The integrations are  over imaginary time and all the vectors within a unit cell.
Therefore, the contribution to $\Gamma$ at second order is
\beq
\Gamma_2 = {1 \over 2} (\Gamma_2^b -\Gamma_2^a) + (\Gamma_2^c -\Gamma_2^{cr})
\eeq

If we consider the  case where the coordinates ${\bf u},{\bf r}$ scan the entire space i.e., $K=1$ and $A=G(x,y)$, we
 find
 that the structure of $\Pi$ yields
 $\Gamma_2^{cr}= \Gamma_2^c$, hence  canceling the contribution of the 2PR
diagram of Fig.2c. Since such cancellations 
occur at all orders, 
the rules (1-3) imply that only two particle irreducible (2PI)
are retained in the expansion for $\Gamma$.  In addition, 
(\ref{joa}) implies that $G_0= G$ and $J_0= G_n^{-1} - G^{-1}$, where $G_n$ is the
non-interacting Green's function in the absence of the source.  Replacing $G_0$ by $G$ in the above
terms, we recover the
Baym Kadanoff functional 
 for the Green's function $G$
\beq
\Gamma[G]= {\rm Tr}\log G - {\rm Tr} G_n^{-1} G+ \Gamma_1 + I_2[G]
\label{bk}
\eeq
\noindent
where $I_2$ is the sum of all connected two particle irreducible diagrams occurring
at second  and  higher orders.

\subsection{generalization}

The formalism presented above can be used to construct the 
effective action functional for any arbitrary   choice of the operator
$O$ or equivalently, the kernel $K$ in (\ref{oper}). The rules (1-3)
can be extended directly,  
provided $K$ satisfies the following conditions:
\begin{enumerate}
\item  $K({\bf r},{\bf r}^\prime\vert {\bf x},{\bf y})$ is invertible

\item 
$\int K^{-1} ({\bf r},{\bf r}^\prime\vert {\bf x},{\bf y}) 
 K ({\bf r},{\bf r}^\prime\vert {\bf w},{\bf z}) d{\bf r} d{\bf r}^\prime = \delta_{{\bf w}{\bf x}} \delta_{ {\bf y}{\bf z}}
$
\item 
$\int K^{-1} ({\bf r},{\bf r}^\prime\vert {\bf x},{\bf y}) 
 K ({\bf u},{\bf u}^\prime\vert {\bf  x},{\bf y}) d{\bf x} d{\bf y} = \delta_{{\bf r}{\bf u}} \delta_{{\bf r}^\prime {\bf u}^\prime }
$
\end{enumerate}
\noindent
We can then use the rules (1-3) to construct $\Gamma[A]$ with the
following replacements  
\beq
\nonumber
G_{0} (R+r ,R^\prime +r ^\prime) = \left[\left(\partial_\tau-
{{\bigtriangledown^{2}}\over {2m}} +
v_{X}({\bf R}+{\bf r})\right)\delta_{{\bf R} {\bf R}^\prime}
\delta(r-r^\prime) + \int d{\bf u} d{\bf u}^\prime
 J({\bf u},{\bf u}^\prime,\tau,\tau^\prime) K({\bf u},{\bf u}^\prime\vert
{\bf R}+{\bf r},{\bf R}^\prime+{\bf r}^\prime)
\right]^{-1}
\eeq
\noindent
and
\bea
\nonumber
\Pi (r,r^\prime,u,u^\prime)  &=& 
\sum_{{\bf R},{\bf R}^{\prime},{\bf U}, {\bf U}^\prime}\int_{{\bf a},{\bf a}^\prime,{\bf b},{\bf b}^\prime} 
K({\bf r},{\bf r}^\prime \vert {\bf R}+{\bf a},{\bf R}^\prime+{\bf a}^\prime) K({\bf u},{\bf u}^\prime\vert
{\bf U}+{\bf b},{\bf U}^\prime+{\bf b}^\prime)
\langle \psi^{+}(R+a)\psi(R^{\prime}
+a^{\prime}) \psi^+(U +b)
\psi(U^\prime +b^\prime) \rangle \nonumber \\
&=& 
\sum_{{\bf R},{\bf R}^{\prime},{\bf U}, {\bf U}^\prime}\int_{{\bf a},{\bf a}^\prime,{\bf b},{\bf b}^\prime} 
K({\bf r},{\bf r}^\prime \vert {\bf R}+{\bf a},{\bf R}^\prime+{\bf a}^\prime) K({\bf u},{\bf u}^\prime\vert
{\bf U}+{\bf b},{\bf U}^\prime+{\bf b}^\prime)
G_0(R+a,U^{\prime} +b^
{\prime}) G_0(R^\prime+a^\prime,U +b)
\end{eqnarray}
\noindent
The generalization to the spin dependent case is straightforward.

\section{ Approximations}

We now  address the question of how to calculate $\Gamma_{xc}$ in (\ref{j0sc}). As is the case in the usual density functional theory, 
the   evaluation of  all the diagrams that contribute to $\Gamma$ and
hence $\Gamma_{xc}$ is impossible and one has to take recourse to various
approximation schemes depending on the strength of the interactions. 

\subsection{Weak Coupling} 

The simplest scheme would be  to use truncated perturbation theory
to second order, utilizing the expressions given in the previous section.
Alternatively, one could  use a random phase approximation.
This approximation leads to a simplification of the rules for the
effective action presented above. Since all RPA diagrams are by
definition irreducible with respect to the coulomb vertex and two particle
irreducible, we do not have to go through the complicated procedure  outlined
in rules (3a)-(3e) to
deal with reducible diagrams.
It would be interesting to investigate the equivalent of the 
 local density
approximation (LDA) for these new functionals, which corresponds
to evaluating the  diagrammatic expansion  for $\Gamma_{xc}$ described in the
previous section for a spectral function which is independent
of the spatial coordinates i.e., spatially homogeneous.  Unfortunately,
 the energy of the uniform electron gas
 expressed as a functional of its spectral function is not available  and 
further work is required before a correspondence with LDA can be made.

\subsection{Strong Coupling  - Impurity models}

The weak coupling approximations for $\Gamma$ can be set up for
an arbitrary kernel $K$ in (\ref{oper}).
On the other hand,  different strong coupling
approximations can be set up for different   choices 
 of the  kernel  and are closely related
to 
the formalism of dynamical mean field theory (DMFT) on the lattice 
\cite{review}, where the local Greens function is represented
as a Greens function of a generalized Anderson impurity model.
In finite dimensions, the fact that the exact Green's function
can be represented in terms of a local impurity model is an
assumption. It suggests however, the implementation of 
the local Green's function defined by the kernel
in (\ref{kernel1}). 
For simplicity, we consider the paramagnetic case in the following. 
Denoting the unit cell by  B we represent the local Greens function
defined by the kernel (\ref{kernel1}), 
 as a
{\it generalized impurity model} described by an action
\begin{eqnarray}
S\{G_{o}\} & = & \int d\tau d\tau^\prime  \int_{B} d{\bf r} \int_{B}d{\bf 
r}^\prime \sum_{\alpha} \psi^+_{\alpha} (r)
A^{-1}_{o} ( r,r^\prime)   \psi_{\alpha}(r^\prime
)
\nonumber \\
& + & {\int_{o}}^{\beta}  d\tau d\tau^\prime\int_{B}  d{\bf r} \int_{B}  d{\bf r}^\prime
 \rho( r)
V_{SR}{( r, r^\prime)} \rho (r^\prime)
\label{sgo}
\end{eqnarray}
\noindent
Here $r\equiv({\bf r}, \tau)$ and the density  $\rho(r) = \sum_{\alpha} \psi^+_{\alpha}(r)
 \psi_{\alpha}
(r)$. 
The interaction in the full continuum model is now replaced by an effective
short ranged interaction   $V_{SR}(r,r^\prime)= V_{SR}({\bf r},{\bf r}^\prime) \delta(\tau -\tau^\prime)$  within the cell $B$.
For lattice models, such a mapping led one from the full Green's function
 $G_{ij}$ to
a local Green's function $G_{ii}$.
In the continuum, the mapping leads to a similar "local"
Green's function $A(r,r^\prime)$ which is the Green's function
defined within the unit cell. Referring back to Sec.II,
we see that  a  simple generalization of DMFT ideas
 automatically leads us to consider (\ref{kernel1})
as the simplest   choice  
for local Green's function in the continuum.
Analogy with DMFT on a  lattice, leads to write a self consistent
equation for  $A_o$   by
\beq
A^{-1}_o(r,r^\prime) = [\frac{\partial}{\partial  \tau} -
{{\bigtriangledown^2} \over {2m}}- \mu - V_{X}( r) -V_H(r)]
\delta(\tau - \tau^\prime)\delta({\bf r}-{\bf r}^\prime) + \Delta( r,r^\prime)
\label{go}
\eeq
\noindent
where  $V_H$ is the Hartree potential that arises
naturally in such an impurity mapping
and  $\Delta(r,r^\prime)$ is the  time-dependent
"Weiss field of the problem".
 It describes 
the effect of all the other unit cells  (which have been integrated out),
on the
selected unit cell.  Note that this  looks like a generalization of
the multi-orbital impurity models if we assume that the  orbital indices
are the   continuous
variables ${\bf r}, {\bf r}^\prime$.  
Using $S(A_{o})$ given in (\ref{sgo}) we can compute the "local 
Green's function"  $A(r,r^\prime)$ of the
unit cell defined by 
\beq
\frac{\int d\psi^{+} d\psi  \psi( r)
\psi( r^\prime) e^{S({A_{o}})}}{\int
e^{S\{A_{o}\}}} = A(r,r^\prime)\{\Delta\}
\eeq
\noindent
Note that $A$ depends on the Weiss field $\Delta$. The self
 energy   $ \Sigma[ \Delta ] =A^{-1}_o- A^{-1}(r,r')$  
can be thought of as  the sum of the Hartree term plus all the
two  particle irreducible skeleton graphs with 
interaction  lines $V_{SR}(r,r')$ and particle lines  given by the
"local" 
Green's function 
$A(r,r^\prime)$.  Using the above, the self consistency condition for  
$\Delta(r,r^\prime)$ is obtained by requiring  that
\begin{equation}
A^{-1}(r,r^\prime)[\Delta]=[\frac{\partial}{\partial \tau} -
\frac{\nabla ^{2}}{2m} -\mu - V_{X}(r) - V_{H}(r)[\Delta]] 
\delta({\bf r}-{\bf r}^\prime)\delta(\tau-\tau^\prime) +\Delta(r,r^\prime)
- \Sigma(r,r^\prime)[\Delta]
\label{gimp}
\eeq    
\noindent
This procedure of mapping the model onto an impurity model with interactions defined
only within the unit cell is nothing but an approximation to the infinite series
of diagrams in the perturbative expansion described earlier. Diagrammatically, the present   method  
retains only local diagrams i.e., those whose internal propagators and
vertices just connect 
points within the same  unit cell. None of the diagrams include particle
lines going from one unit cell to another. This is
indeed reminiscent of the  Hubbard model  calculations in infinite dimensions where
the self energy is completely local in space.  
A solution of this self-consistent equation for $\Delta$ is sufficient to
specify the local $A$ and hence other dynamical quantities. 

The above mapping can be   generalized to  
 a more general impurity model which goes under the name
of extended DMFT \cite{si,edmft}.  This captures some important aspects  of the
 long range part of the Coulomb interaction \cite{edmft}.
In the case of lattice models, one obtains a  single impurity 
whose on-site interaction  is dynamically  and self-consistently
screened due to the
presence of long-ranged coulomb interactions in the original lattice
model.
This has been worked out explicitly for the lattice model in
 Ref.\onlinecite{edmft}. A straight forward generalization 
 to the continuum is obtained by    replacing the  
 short-ranged interaction $V_{SR}(r,r^\prime)$ in (\ref{sgo}),
by the 
 retarded time dependent  interaction 
\beq
\Pi_o^{-1}(r,r^\prime)= V_{SR}({\bf r},{\bf r}^\prime)\delta(\tau-\tau^\prime)
+\Omega(r,r^\prime) 
\label{dyimp}
\eeq
In the same manner as above, the local
density density correlation function 
 $\Pi= \langle \rho(r) \rho(r^\prime) \rangle$
is given by
\begin{equation}
 \Pi^{-1}(r,r^\prime)[\Omega,\Delta] = \Pi_o^{-1}-{\tilde \Pi}[{\Omega,\Delta}]
\label{den}
\end{equation} 
\noindent
where $\tilde \Pi$ is the irreducible part of  $\Pi$.   
We, therefore, see that since $\Pi$ depends on the retarded correlations, the dynamical
interaction $\Pi_o$ is also determined in a self-consistent manner.
Moreover, in the presence of this retarded interaction, the self energy $\Sigma = \Sigma[\Delta,\Omega]$ i.e., it depends on $\Delta$ through $G_o$ and also
on $\Omega$  through
the interaction  $\Pi_o$. Therefore, (\ref{gimp}) and (\ref{den})
form a set of coupled equations for
$\Delta$ and $\Omega$  which have to be solved self-consistently. These are
the  generalizations of the extended DMFT equations obtained in Ref.\onlinecite{edmft}.
 The diagrammatic content of this approach is slightly different from that 
of the unscreened model since  the vertex $V_{SR}$  is now replaced by $\Pi_o$.
 This implies that in the full continuum model, we retain all those "local"
 diagrams  with  internal particle lines $A(r,r^\prime)$  and long ranged
vertices  which   connect any 
pair of arbitrary points ${\bf r}+{\bf R}$ and ${\bf R}^\prime +{\bf r}^\prime$.
 
To summarize, we see that  DMFT ideas can be generalized to provide 
strong coupling approximations
 to the  finite dimensional effective actions of the local Green's function $A$.
We briefly mention that the above mappings can be generalized to 
include the effects of  magnetic 
ordering.
In this paper, we have presented the generalized impurity model only for the
simplest kernel and it would be interesting to study whether there exist
 other  generalized impurity models
for arbitrary choices of the local Green's functions.
In particular we notice that current DMFT calculations using 
realistic  multiband Hamiltonians \cite{lichtenstein} based
on LMTO orbitals can be viewed as approximate realizations
of the functional of the local Green's function based on
the choice of the kernel (\ref{kerorb}), where the tight binding
basis defining
the kernel are currently  obtained   
from the self consistent solutions of a density functional
calculation.

\section{conclusions} 

In this paper, we have described a means of   studying   electronic structure
problems  defined in the continuum, involving effective action formalisms
and the principles of dynamical mean field theories. 
We first outlined the method to formulate  the effective
action functional for a general "local" Green's function. These
local functions  
 are good probes of the dynamics of the system including metal
insulator transitions.
Though  we present a scheme to calculate these
functionals   to every order in the interaction, as is the case with
standard density functional theory,  an exact evaluation of these
functionals is practically intractable. We have, therefore,
 described various weak and
strong coupling approximations. The strong coupling approximations
are a generalization of the DMFT ideas on a lattice to the continuum. 
Recently, some of these  generalized DMFT ideas  were applied to the
 problem of the Hubbard model with coulomb long range interactions. 
 \cite{edmft}. We found that the long range interactions screen the effective onsite
interaction  in such a manner as to make the continuous Mott
transition seen in that system,  discontinuous. 
These  results suggest that a direct implementation of
DMFT to an electronic structure problem  can produce results which are
not just quantitatively but also 
qualitatively different from  those  obtained from    conventional DMFT 
model calculations.

\end{document}